\documentstyle[pra,aps,epsf,floats]{revtex}

\begin{document}

\draft

\twocolumn[\hsize\textwidth\columnwidth\hsize\csname@twocolumnfalse%
\endcsname

\title{Chaos in the random field Ising model}

\author{Mikko Alava$^{1,2}$ and Heiko Rieger$^{3,4}$}

\address{
$^1$ NORDITA, Blegdamsvej 17, 2100 Copenhagen, Denmark\\
$^2$ Helsinki University of Technology, Laboratory of Physics,
P.O. Box 1100, 02015 HUT, Finland\\
$^3$ Institut f\"ur Theoretische Physik, Universit\"at zu K\"oln, 
     50923 K\"oln, Germany\\
$^4$ HLRZ, Forschungszentrum J\"ulich, 52425 J\"ulich, Germany\\
}

\date{March 18, 1998}

\maketitle

\begin{abstract}
  The sensitivity of the random field Ising model
  to small random perturbations of the quenched disorder
  is studied via exact ground states obtained with a maximum-flow
  algorithm. In one and two space dimensions we find a mild form of
  chaos, meaning that the overlap of the old, unperturbed ground state
  and the new one is smaller than one, but extensive. In
  three dimensions the rearrangements are marginal (concentrated in the
  well defined domain walls). Implications for finite temperature
  variations and experiments are discussed.
\end{abstract}


]

\newcommand{\bc}{\begin{center}}
\newcommand{\ec}{\end{center}}
\newcommand{\be}{\begin{equation}}
\newcommand{\ee}{\end{equation}}
\newcommand{\beqn}{\begin{eqnarray}}
\newcommand{\eeqn}{\end{eqnarray}}

The concept of {\it chaos} in disordered systems refers to the
sensitivity of their equilibrium state (at finite temperatures) or
ground state (at zero temperature) with respect to infinitesimal
perturbations. In spin glasses \cite{review} for instance it is well
known that small changes of parameters like temperature or external
field causes a complete rearrangement of the equilibrium configuration
\cite{fisher,bray}.  This has experimentally observable consequences
like reinitialization of aging in temperature cycling experiments
\cite{tempcyc} and has also been investigated in numerous theoretical 
works \cite{sgtemp}.

A slight random variation of the quenched disorder has the very same
effect on the ground state configurations. Although of
similar origin, chaos with respect to temperature changes is harder to
observe than chaos with respect to disorder changes
\cite{sgcomp}, and the latter phenomena has been used to 
quantify spin glass chaos in numerical investigations \cite{bray,sg0}.

This type of chaos has actually later been discovered in another,
simpler random system, the directed polymer in a random
medium \cite{zhang,natter,feigel}, which is equivalent to an
dommain wall in a random bond ferromagnet. The interface displacement
as a reaction to infinitesimal random changes of bond-strengths obeys
particular scaling laws with exponents related to the well-known
interface roughness exponent $\chi$ \cite{natter,feigel}.

In this paper we consider the random field Ising model \cite{review}
and study, the first time to our knowledge, the sensitivity of its
ground state with respect to small changes in the random field
configurations. It turns out that the emerging picture is very
reminiscent of chaos in spin glasses and random interfaces. 
This statement is quantified by the following
phenomenological picture \cite{bray,natter}.

Consider a random Ising system defined, for instance, by the Hamiltonian
\be
H=-\sum_{\langle ij\rangle} J_{ij} S_i S_j -\sum_i h_i S_i
\ee
where $S_i=\pm1$ are Ising spins, $\langle ij\rangle$ indicates
nearest neighbor pairs on a $D$-dimensional lattice of, say,
linear size $L$, the $J_{ij}$ denote interaction strengths and $h_i$
local fields, both quenched random variables obeying some distribution
(continuous, in order to exclude ground state degeneracies). The case
$J_{ij}$ Gaussian (with mean zero and variance one) and $h_i=0$ is the
{\it spin glass} (SG), the case $J_{ij}\ge0$ and $h_i=0$ is
the {\it random bond ferromagnet} (RBFM) and the case $J_{ij}=J$ and
$h_i$ Gaussian (with mean zero and variance $h_r$) is the {\it random field
Ising model} (RFIM).  In order to study the sensitivity of the ground
state of these systems with respect to small changes in the quenched
disorder we can apply a random perturbation of amplitude $\delta \ll 1$ to 
any of the quenched random variable.  As a consequence the 
{\it  new} ground state will differ from the old one.

The RFIM ground state changes when the
domain structure changes (for purely ferromagnetic states 
this argument does not work). One can estimate when the two ground 
states will be uncorrelated, beyond a length scale $L^*$
This can be found considering domain walls with an Imry-Ma
\cite{imry} type argument \cite{bray,natter}: The energy $E_{flip}$ to
flip droplets or domains or excitations of size $L$ scales like
$L^\theta$, where $\theta$ is the energy fluctuation exponent ($\theta$ is
denoted $y$ in the SG context \cite{bray}). The energy change due to
the random perturbation $E_{rand}$ scales like $\delta L^{d/2}$, where
$d=d_s$ the fractal dimension of the droplet's surface in the SG case,
$d=D-1$ the interface dimension in the RBIM and $d=D$ in the RFIM
case. The decorrelation takes place when $E_{rand}(L)>E_{flip}(L)$, i.e.\ for
$L>L^*\sim\delta^{-1/\lambda}$ with $\lambda=d/2-y$.
In SG jargon $L^*$ is called the overlap length and
$\lambda$ is denoted $\zeta$, the chaos exponent \cite{bray,sg0}.

Two remarks are in order: first, as pointed out in \cite{natter} and
\cite{feigel} already for $L<L^*$ the ground state is slightly altered
by the random perturbation. This is, however, an effect of the
interplay between elastic energy and $E_{rand}$. This leads to
displacements of the domain wall of size $\Delta x\sim\delta
L^{\alpha}$ with $\alpha=\lambda+\chi$, where $\chi$ is the roughness
exponent. The roughness exponent \cite{fisher2} and the energy
fluctuation exponent $\theta$ are related via $\theta=2\chi+D-3$
\cite{natter}.

The second remark concerns the RFIM-case: for $D\le2$ the concept of a
macroscopic domain wall fails here, and the above considerations can
only be transfered {\it cum grano salis}. This means that they are
sensible only for $L\ll\xi\sim\exp(-h_r^2/A)$, the typical size of
domains in the 2D RFIM \cite{binder,seppala}.

In 3D the situation is different: The concept of a domain wall is well
defined and we get with the estimate for the roughness exponent
$\chi=2/3$ \cite{fisher2} and, consequently, $\theta=4/3$ the result
$\lambda=1/6$, i.e.\ $L^*\sim\delta^{-1/6}$.  The typical displacement
of a domain wall thus is, as above, $\Delta x\sim L^\alpha$ with
$\alpha=2/3$ for $L\ll L^*$ and $\alpha=5/6$ for $L\gg L^*$.

\begin{figure}
\epsfxsize=\columnwidth\epsfbox{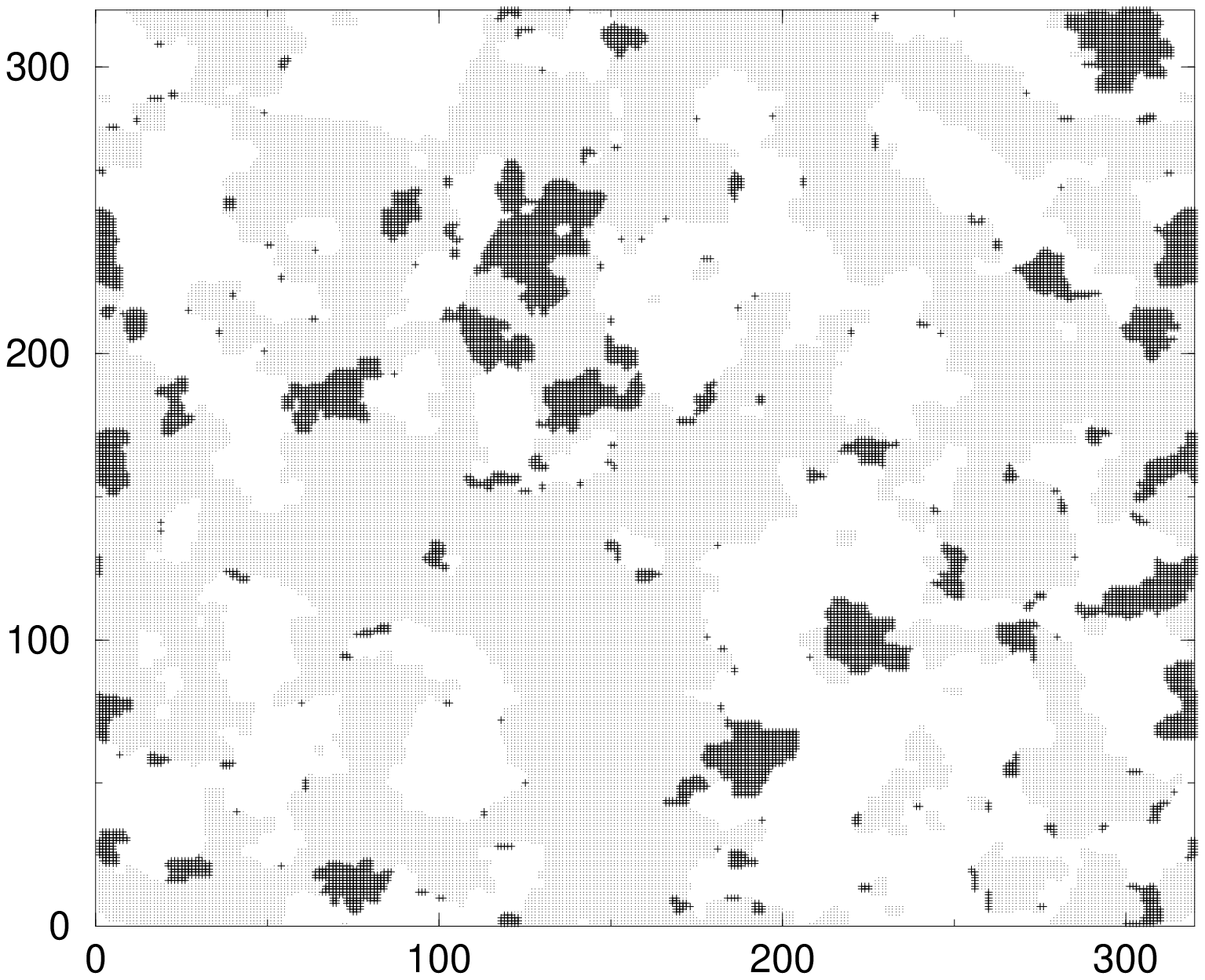}
\caption{
\label{domains}
A ground state plus the perturbation-induced changes.
The original spin-orientations are indicated in grey for $S_i =+1$ and
white for $S_i =-1$, the flipped spins are indicated in black. $L=320$,
$\Delta=2$ and $\delta=0.1$ (see text).}
\end{figure}

If we take the above arguments serious for the RFIM in 2D the overlap
length $L^*$ turns out to be formally infinite (since with $\theta=1$
one has $\lambda=0$), where one has of course to be careful due to
logarithmic corrections to the energy $E_{flip}$.  Thus the mechanism
by which rearrangements take place is due to the interplay between
elastic energy and $E_{rand}$. Moreover, in the 2D RFIM, the typical
displacement of domain walls should scale as $\Delta x\sim\delta L$
for $L\ll\xi$ since $\alpha=\lambda+\chi=1$. As a conseqence the
correlation or overlap between the old, unperturbed, ground state
$S_i$ and the new one $S_i'(\delta)$
\be
q=\frac{1}{L^D}\sum_i S_i S_i'(\delta)
\ee
\noindent
behaves like $1-q\sim L^{\alpha-1}$ and therefore $q$ should be of
order ${\cal O}(1)$, depending on the probability with which domain
wall displacements occur.

In what follows we present results of exact ground state calculations
for 1D spin chains and for 2D systems. We use a random field
distribution and a perturbation distribution that have a constant
probability density between $-\Delta$ and $\Delta$ and $-\delta/2$ and
$\delta/2$, respectively and set $J_{ij}=1$. Figure 1 shows an example
of a large 2D ground state ($L=320$) with the two spin orientations
shown in white and grey, respectively, and the {\it flipped} spins in
black. There are two features one should note.  First, the size of the
system is larger than the critical length scale needed for ground
state breakup and the magnetization is practically zero. Second, the
flipped spins form a number of clusters of varying size, that seem to
concentrate on the {\it cluster boundaries} of the original
groundstate.

\begin{figure}
\epsfxsize=\columnwidth\epsfbox{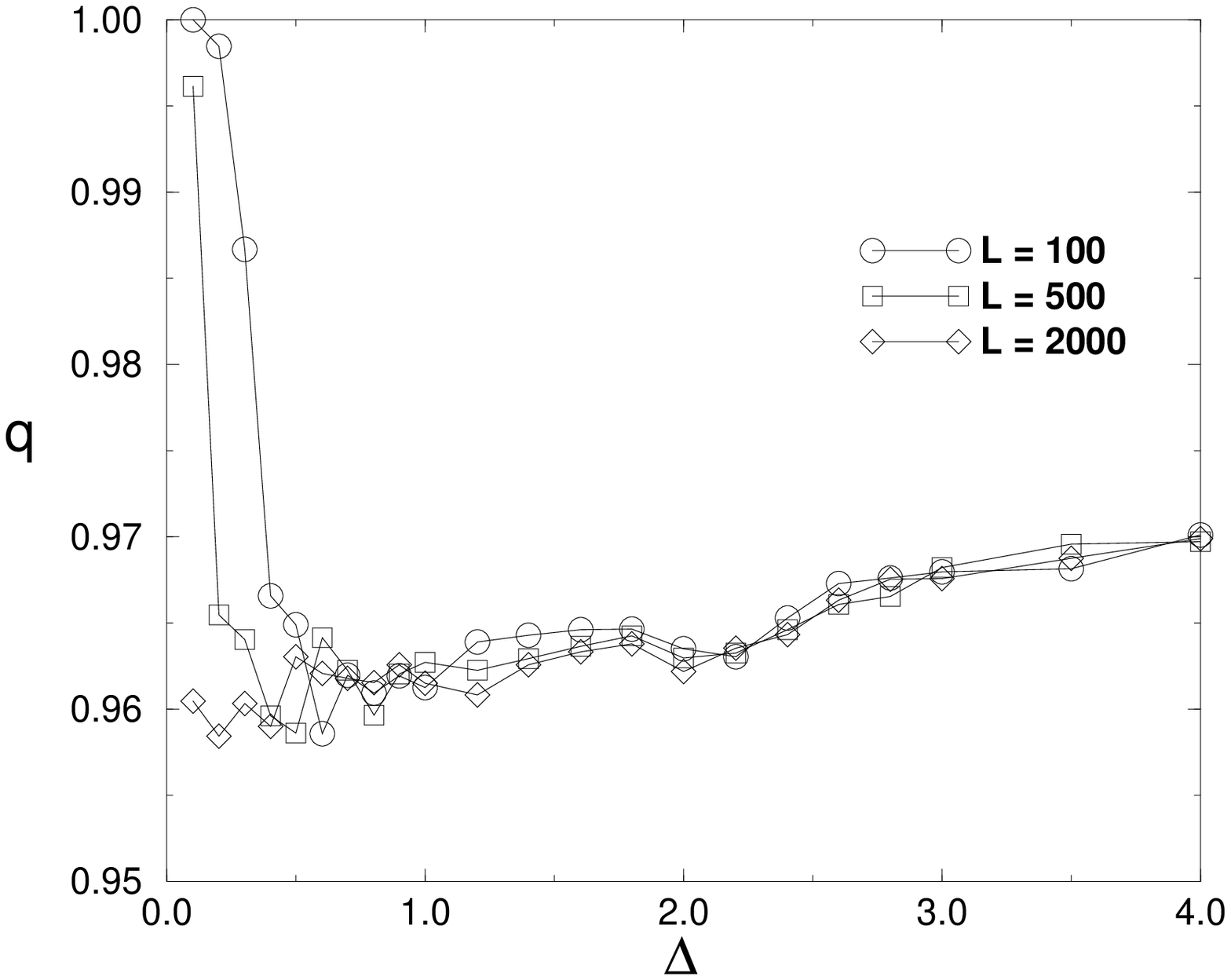}
\epsfxsize=\columnwidth\epsfbox{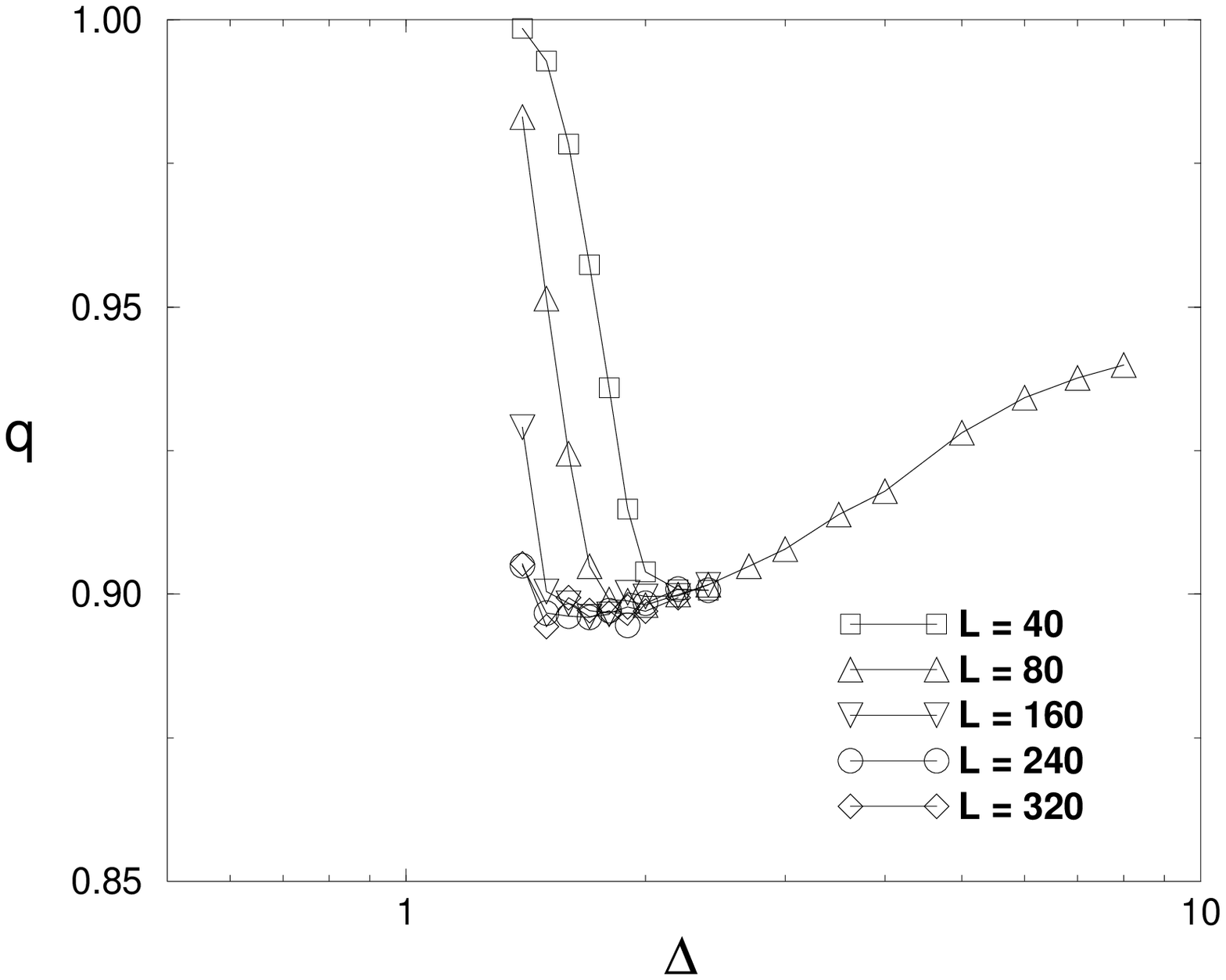}
\caption{
\label{overl}
{\bf a)} (Top) Scaling of the overlap parameter with random field
strength for the 1D spin chain. {\bf b)} (Bottom) Scaling of the
overlap in 2D for the system sizes $L=$ 40,80,160,240 and 320 and for
$\delta=0.1$.}
\end{figure}

Figure \ref{overl} shows what happens as one sweeps the RF-strength
($\Delta$). In arbitrary dimensions, the limit $\Delta \rightarrow
\infty$ goes over to a site percolation problem, i.e. the local
RF-orientation gives the spin state at a site. In that limit the
overlap $q$ is determined by the probability of the applied
perturbation $\delta$ to change the this orientation. For somewhat
smaller fields $q$ gets smaller and in an apparently linear fashion as
$\Delta$ changes. In the 1D case the overlap is not sensitive to the
system size above a certain threshold in $\Delta$, below which the
overlap quickly increases to unity again, which indicates a typical
domain size. The overlap seems to become a $\delta$-dependent constant
in the thermodynamic limit and for $\Delta \rightarrow$ 0.

\begin{figure}
\epsfxsize=\columnwidth\epsfbox{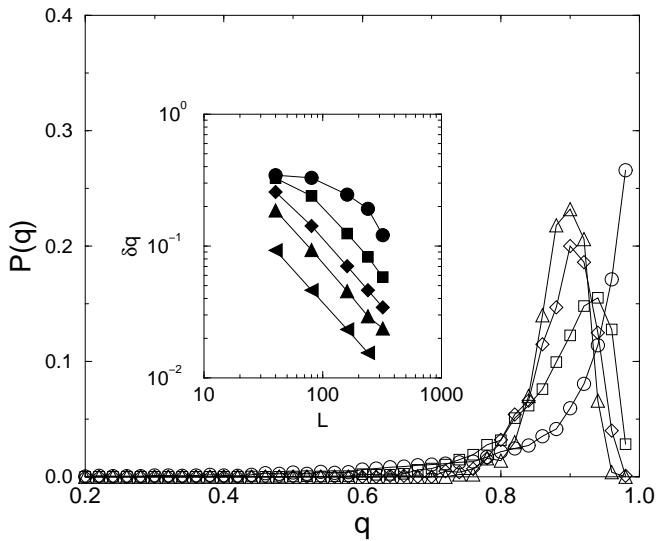}
\caption{
\label{q-pdf}
Probability distributions of the overlap $q$ for $\Delta =1.8$ for
the system sizes $L= 40 \dots 320$. The inset shows the standard deviations
of the overlap pdf's for $\Delta = 1.4, 1.6, 1.8, 2, 2.4$. }
\end{figure}

This 1D behavior can be understood as follows. For simplicity let us
assume that the first spin is fixed to be up, i.e.  $S_0=+1$. Then the
total random field energy at site $n$ is given by $H_r=\sum_{i=1}^n
h_i$ in the unperturbed system and $H'_n=H_n+\Delta_n$ with
$\Delta_n=\sum_{i=1}^n\delta_i$, in the perturbed one. If $h_i$ and
$\delta_i$ are independently distributed variables with zero mean and
variance $h_r=[h_i^2]_{\rm av}$ and $\delta_r=[\delta_i^2]_{\rm av}$,
respectively, the variables $H_n$ and $\Delta_n$ are (for $n\gg1$)
Gaussian with mean zero and variance $nh_r$ and $n\delta_r$,
respectively. The probability distribution $P(H_n,H'_n)$ is simply
given by
$$
P(H_n,H'_n) = \int d\Delta_r P(H_n) P(\Delta_n)
\delta(H_n+\Delta_n-H'_n)\;.
$$
Now the total RF fluctuations $H_n$ and $H_n'$ produce domains if
their magnitude is large enough to overcome the ferromagnetic
coupling: suppose that $S_i=+1$ and $H_i>-J$ for $i=1,\ldots n$ (i.e.
a plus-domain), but $H_{n+1}<-J$; then $S_{n+1}$ will be 
flipped, i.e.\ $S_{n+1}=-1$ and a new (minus-) domain 
starts. For large enough typical domain sizes the total RF
fluctuations become large: one can neglect $J$ and assume that only
the sign of $H_n$ and $H_n'$ determines the ground state (note that this
is different from the high field region $h_n\gg J$, in which the local
random fields $h_i$ dominate).Thus, the probability
for $S_n$ and $S_n'$ being equal is given by
\be
p(S_n=S_n') = 
\int dH_n\,dH_n'\,P(H_n,H'_n)\cdot\theta(H_n\cdot H'_n)\;,
\ee
where $\theta$ is the step function. A straightforward calculation
yields $p(S_n=S_n')=1-\frac{1}{\pi}\delta_r/h_r+{\cal O}(\delta^2)$.
For the data shown in Fig. 1a, in which $h_r^2=\Delta^2/3$ and
$\delta_r^2=\delta^2\Delta^2/12$ with $\delta=0.1$, we have
$\delta_r/h_r=0.05$ and hence $q=-1+2p(S_r=S_r')\approx0.97$ agreeing
roughly with the numerical results for $h_r\to0$ in the limit
$L\to\infty$.

\begin{figure}
\epsfxsize=\columnwidth\epsfbox{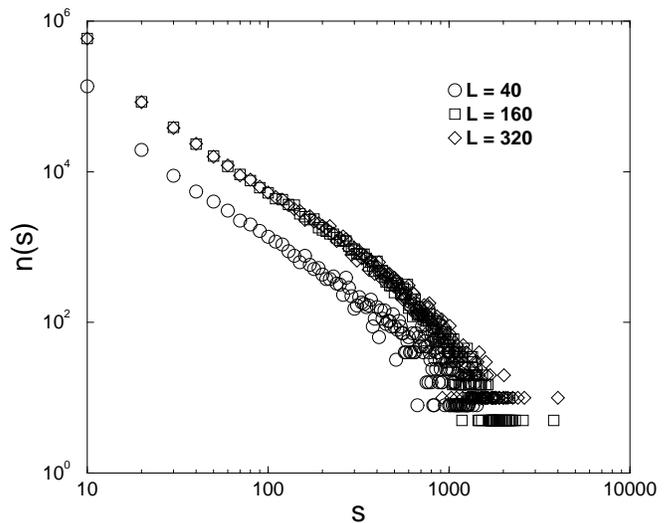}
\caption{
\label{clusdist}
Cluster size distributions of clusters of flipped spins for
$\delta=0.1$, $L=$ 40, 160, 320, and $\Delta= 2$.}
\end{figure}

The 2D behavior is depicted in Figure \ref{overl} for $\delta =0.1$. 
The number of simulations is 10000 for $L=$ 40 and 80, 4000 for $L=$ 160,
1000 for $L=$ 240 and 500 for $L= 320$. The generic behavior of the 
overlap is as for the 1D chain:  $q(\Delta)$ is
roughly linear until the regime of small fields ($\Delta \leq 2$)
after which it seems to saturate to a $\delta$-dependent value
$q(\delta )$. The cross-overs (increase of $q$ with decreasing
$\Delta$) are due to the ground state breakup mechanism. For
small systems the ground state is ferromagnetic, except for 
a limited number of domains of the opposite spin orientation.
The decrease in $q$ is caused by the effect of the ground state
becoming more and more uniform (magnetization $|m| \rightarrow 1$).
Otherwise the behavior resembles strongly the 1D case.

The thermodynamic behavior of the overlap is also visible in the
statistics of overlap distributions. Figure \ref{q-pdf} shows how the
probability distribution $P(q)$ of $q$ behaves with varying system
size and for $\Delta = 1.8$ (the data is the same as presented in Fig.
\ref{overl}). For all systems $P(q)$ is peaked at $q=1$ but as $L$
is increased a peak appears in the distribution, resembling a
Gaussian. The inset of Fig.~\ref{q-pdf} shows the standard deviation
$\delta q$ of $P(q)$ for varying $\Delta$ as a function of the system
size $L$. Except for the by-now standard cross-over for small
$L$ and $\Delta$ we observe, that the width of the distribution
{\it decreases}, which signals that in the thermodynamic limit 
$P(q)$ approaches a delta-function-like sharp
one. The cross-over exponent $c$, defined with $\delta q \sim L^{-c}$,
seems to be exactly one ($c=1$).

The mechanism by which $q$ is determined is illustrated in
Figure \ref{clusdist}. The size distribution of flipped
clusters $n(s)$ converges with $L$ to a power-law, $n \sim s^{-1.6}$,
with a cut-off that depends very weakly if at all
on $L$. This has to be so for
the overlap not to diverge to zero in the thermodynamic limit,
since one can write $1-q$ as an integral over $n(s)$: a $L$-dependent
cut-off would imply that $q$ would decrease continuously.

\begin{figure}
\epsfxsize=\columnwidth\epsfbox{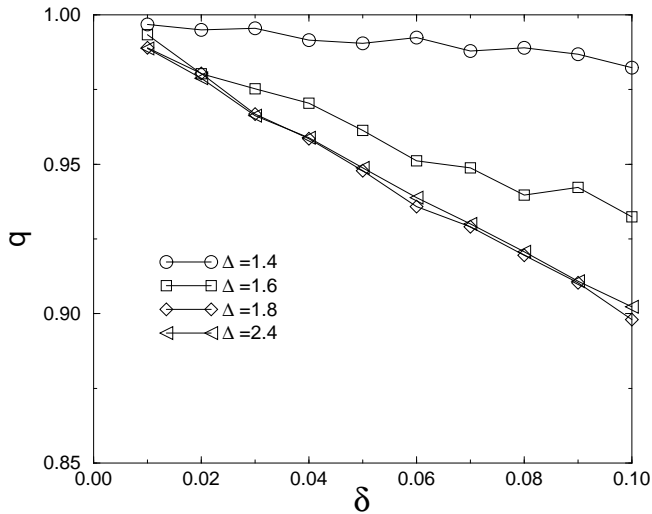}
\caption{
\label{deltadep}
Dependence of the overlap $q$ on the perturbation strength $\delta$
for weak and strong magnitudes ($\Delta$),  $L=80$.}
\end{figure}

Finally in Figure \ref{deltadep} we demonstrate that $1-q \sim \delta$
for small $\delta$. This follows from the scaling arguments presented
for 2D RFIM domain walls and the 1D RF chain.

In this paper we have considered the stability of the random field
Ising model to small perturbations. Unlike in spin glasses, it turns
out that the RFIM ground state shows a weak form of chaos, similar to
directed polymers or random bond Ising model domain walls. The overlap
$q$ attains its value from fluctuations of domain walls, in both 1D
and 2D. Thus the ground state stays almost intact.  The ground state
domains are robust against external perturbations since, most likely,
the field excess of a domain is extensive ($\sum h_i \sim V$). For the
RFIM in 3D the prediction of the domain wall scaling argument is that
$q$ should converge to unity since the domain wall displacement
exponent $\alpha$ here is $5/6$: the displacement of a domain wall on
large enough length scales is $\Delta x\sim L^\alpha$ and therefore
$1-q\propto L^{\alpha-1}\to0$. Moreover, in both limits $h_r/J\to0$
and $h_r/J\to\infty$, i.e.\ deep in the ferromagnetic phase and deep
in the paramagnetic phase, it is trivial that $q\to1$.

One would like to extend the argumentation
to changes in temperature as is common for spin glasses and
random bond -type directed polymers. In spin glasses 
chaos is intimately linked to the non-equilibrium
correlation length, which gives rise
to measurable consequences in e.g. temperature cycling
experiments that measure the out-of-phase susceptibility.
Here, however, repeating the
scaling argument of domain wall for temperature changes
results in a displacement exponent which does not produce
any extensive changes in the overlap. In 2D the predicted
outcome is simply that of a random walk ($\Delta x \sim L^{1/2}$).
In other words assuming that typical valleys in the energy 
landscape are separated by an energy given by the energy
fluctuation exponent gives completely different results
for temperature and ground state chaos than for random
bond disorder. This discussion is intimately related
to coarsening and ageing in the RFIM;
one should note that there are so far no simulation results
that address these questions directly.

\acknowledgements

This work has been performed within the Finnish-German cooperation
project supported by the Academy of Finland and the DAAD.
M.A. would like to thank Eira Sepp\"al\"a for a version
of the computer code used.

\end{document}